\documentclass[twocolumn]{webofc}
\usepackage[varg]{txfonts}   
\usepackage{amsmath}
\begin{document}
\title{Isospin Conservation in Neutron Rich Systems of Heavy Nuclei}
%
%
\author{\firstname{Ashok Kumar} \lastname{Jain}\inst{1}\fnsep\thanks{\email{ajainfph@iitr.ac.in}} \and
        \firstname{Swati} \lastname{Garg}\inst{1}\fnsep}

\institute{Department of Physics, Indian Institute of Technology Roorkee, Roorkee-247667, India}

\abstract{%
It is generally believed that isospin would diminish in its importance as we go towards heavy mass region due to isospin mixing caused by the growing Coulomb forces. However, it was realized quite early that isospin could become an important and useful quantum number for all nuclei including heavy nuclei due to neutron richness of the systems~\cite{robson}. Lane and Soper~\cite{lane} also showed in a theoretical calculation that isospin indeed remains quite good in heavy mass neutron rich systems. In this paper, we present isospin based calculations~\cite{jain,swati} for the fission fragment distributions obtained from heavy-ion fusion fission reactions. We discuss in detail the procedure adopted to assign the isospin values and the role of neutron multiplicity data in obtaining the total fission fragment distributions. We show that the observed fragment distributions can be reproduced rather reasonably well by the calculations based on the idea of conservation of isospin. This is a direct experimental evidence of the validity of isospin in heavy nuclei, which arises largely due to the neutron-rich nature of heavy nuclei and their fragments. This result may eventually become useful for the theories of nuclear fission and also in other practical applications.
}
\maketitle
\section{Introduction}
\label{intro}
Isospin is a useful and fundamental quantum number in nuclear and particle physics. In nuclei, we consider neutron and proton to be the two isospin states of a common entity called a nucleon with isospin $T=1/2$. The third component of isospin is different for both the states, $T_3=+1/2$ for neutron and $T_3=-1/2$ for proton. This convention is just opposite to what is normally used in particle physics. 

\begin{figure}[h]
\centering
\includegraphics[width=10cm]{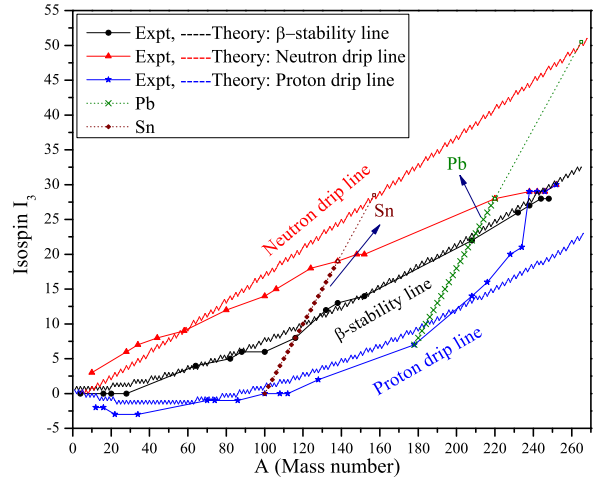}
\caption{Experimental neutron-rich nuclei (upper solid line), $\beta$-stable nuclei (mid-line) and neutron deficient nuclei (lower solid line), plotted from the known  experimental data. The wavy lines show the theoretical predictions for the drip line nuclei.}
\label{driplines}
\end{figure}

We plot a graph similar to nuclear landscape in Fig.~\ref{driplines}, but in terms of isospin $T_3$ on one axis. We note that $T_3=(N-Z/2)$ for a given nucleus and is a well defined quantity. The line in the middle (black in color) shows the line of $\beta$-stability. We have the neutron rich nuclei on the left of the line of stability and the neutron deficient nuclei on the right, plotted on the basis of the presently known experimental data. Theoretical predictions for the drip lines are shown by the wavy lines, while the presently known experimental limits are shown by joining the known data points. We can see that experimentally known neutron-rich nuclei are quite far from the theoretical predictions. On the other hand, the experimental data touch the theoretical predictions for the proton drip line at least up to $A=200$. Around  $A=240$, the experimentally known neutron rich and neutron deficient nuclei merge into the near-stability line, suggesting that a large number of neutron-rich and neutron-deficient isotopes are yet to be found.

There are two main features in Fig.~\ref{driplines} that we would like to emphasize. Firstly, as we go from light to heavy nuclei, the isospin increases due to the fact that $N>Z$ in heavier nuclei. Secondly, in a given chain of isotopes, the range of isospin values could be very large. For example, in the chain of Pb isotopes,  the isospin values could range from near zero in the neutron deficient isotopes to nearly 50 in the neutron rich isotopes. In the most stable lead isotope $^{208}$Pb, the ground state isospin would be $T=22$ as $T_3=22$. It is thus obvious that very large $T$ values are involved in dealing with heavy nuclei. These two factors make the heavy nuclei very interesting to study. We also note that the fission fragments coming out from such sources will be even more neutron-rich. Therefore, the experimental data on fission fragment distributions of heavy nuclei is a good testing ground for verifying the conservation of isospin in heavy nuclei.

However, the data where this could be directly tested is very scarce. Fission fragment distribution data are available in plenty but the mass resolution is about 3-4 units. This means that the $Z$ and $A$ of each fragment is not known precisely in most of the cases. In many situations, particularly the thermal neutron fission, the fragment mass distribution of heavy fragments is still not known. Only recently, more precise fragment distribution data of HI induced fission are becoming available where gamma ray spectroscopy of fragments are being used to identify each fragment, although only in even-even nuclei so far. This in itself presents a huge opportunity for experimnetalists so that good fragment distribution data become available for all the partitions and also for odd-$A$ fragments eventually.

For the first time in 1962, Lane and Soper~\cite{lane} theoretically demonstrated that isospin may remain as a good quantum number in heavy nuclei as in light nuclei. In simple terms, if we consider a nucleus having $N$ neutron and $Z$ protons, we may look upon it as made up of a core ($N=Z$) nucleus having isospin zero and ($N-Z$) valence neutrons. These excess neutrons act in a way so as to reduce the isospin impurity in the nucleus. As the number of excess neutrons rises, the isospin tends to become more pure quantum number. Sliv and Kharitonov (1965)~\cite{sliv} calculated the isospin admixture in light ($N=Z$) nuclei and heavy ($N>Z$) nuclei by using harmonic oscillator shell model wave functions and showed that the isospin admixture in the ground state of $^{16}$O is nearly same as in $^{208}$Pb. A detailed discussion of some these developments may be found in the review by Auerbach~\cite{auerbach}.

In the backdrop of this discussion, we analyze the heavy ion induced fission-fusion reaction $^{208}$Pb($^{18}$O, f) in which neutron-rich fission fragments are emitted. We treat the isospin  to be a conserved quantity and follow the isospin algebra. We calculate the fission fragment mass distribution using this concept of goodness of isospin. An important conjecture given by Kelson~\cite{kelson} is quite helpful in this respect. Kelson has given a theoretical explanation of how n-emission in fission leads to the formation of Isobaric Analog states (IAS) in final fission products. Kelson’s ideas help in resolving our problem of assigning isospin values to various fission fragments. We find that our calculated values match experimental data reasonably well. There are a some deviations which may be due to the presence of shell closure or the presence of isomers.

\section{Formalism}
\label{sec-1}
Consider a projectile $Y$ with isospin $T_Y=T_{3_{Y}}$ incident upon a target with $T_X=T_{3_{X}}$, leading to a compound nucleus ($CN$) with $T_{CN}=\mid T_X-T_Y \mid,......,T_X+T_Y$. However, from isospin algebra which behaves quite similar to SU(2) algebra of spin while dealing with neutrons and protons, $T_{CN} \geqslant T_{3_{CN}}$ and $T_{3_{CN}}=T_X+T_Y$. We assume that only ground or, low lying states of target and projectile are involved. Therefore, the $CN$ always has only one possible value of isospin, $T_{CN}= T_{3_{CN}}=T_X+T_Y$. For the reaction under consideration in present work $^{208}$Pb ($^{18}$O, f), we have $T_X(^{208}Pb)=T_{3_{X}}=22$ and $T_Y(^{18}O)=T_{3_{Y}}=1$. The isospin of $CN$ ($^{226}$Th) can have values $T_{CN}=21, 22$ or 23 but since $T_{3_{CN}} =23$, there is only one allowed value of $T_{CN}=23$. This CN further fissions to give two fragments $F_1$ ($T_{F1},T_{3_{F1}}$) and $F_2$ ($T_{F2},T_{3_{F2}}$) with the emission of $n$ number of neutrons. This is many body problem and to simplify it to two-body problem, we invoke the concept of residual compound nucleus ($RCN$) formed after the emission of $n$ number of neutrons and has a isospin $T_{RCN}$. Now, our complete reaction will look like,
\begin{eqnarray*}
Y(T_Y,T_{3_{Y}})+X(T_X,T_{3_{X}}) \rightarrow CN(T_{CN},T_{3_{CN}}) \\
\rightarrow RCN(T_{RCN},T_{3_{RCN}}) +n \\
\rightarrow F_1(T_{F1},T_{3_{F1}})+F_2(T_{F2},T_{3_{F2}})+n
\end{eqnarray*}

and isospin of $RCN$ should satisfy the two conditions,
\begin{eqnarray*}
\begin{split}
\mid T_{CN}-n/2 \mid \leq T_{RCN} \leq (T_{CN}+n/2)\\
\textrm{and} \quad \mid T_{F1}-T_{F2} \mid \leq T_{RCN} \leq (T_{F1}+T_{F2})
\end{split}
\end{eqnarray*}

Now, we would like to assign isospin values to various fission fragments emitted in different partitions, which is not so straight forward. We formulate two conjectures based on the ideas put forth by Kelson~\cite{kelson} in assigning the isospin values to fission fragments. First conjecture states that neutron emission from $CN$ leads to formation of highly excited states with $T >T_3$. Using this conjecture, we fix the isospin of $RCN$, $T_{RCN}=T_{F1}+T_{F2}$. Kelson’s second conjecture states that the fission fragments are preferably emitted in isobaric analog states (IAS). We thus assign isospin to fission fragments on the basis of this second conjecture. We make three isobars of each mass number having $T_3$ values as $T_3$, $T_3-2$ and $T_3-4$. Then, we assign $T=T_3$ to that particular mass number since this is the minimum value of isospin required to generate all the members of isobaric multiplet formed corresponding to that mass number. For $^{208}$Pb($^{18}$O, f), we have six experimentally observed partitions. We consider eight nuclides on the lighter side and eight on the heavier side of a partition and for each mass number, we make three isobars. Now, we assign isospin to each mass number using Kelson’s second conjecture as described above. For example, for $A=112$, we have three isobars namely $^{112}$Ru with $T_3=12$, $^{112}$Pd with $T_3=10$ and $^{112}$Cd with $T_3=8$. We assign maximum of three $T_3$ values which is 12 to $A=112$ as it can generate all the three isobars $^{112}$Ru, $^{112}$Pd and $^{112}$Pd. 

Once we have assigned the isospin values, we proceed to calculate the relative yields of fragments by using isospin part of the wave function. For a particular pair of fragments emitted in a given $n$-emission channel, the isospin wave function for $RCN$ can be written as,
\begin{eqnarray}
\begin{split}
\mid{T_{RCN},T_{3RCN}}\rangle_n = \langle{T_{F1}T_{F2}T_{3_{F1}}T_{3_{F2}} \mid T_{RCN}T_{3_{RCN}}}\rangle \\ \mid{T_{F1},T_{3_{F1}}}\rangle \mid{T_{F2},T_{3_{F2}}}\rangle
\end{split}
\end{eqnarray}

The first part of the equation on R.H.S. gives us the isospin C.G. coefficient (CGC). The intensity can be calculated by taking the square of CGC,
\begin{equation}
I_n = \langle{CGC}\rangle^2 = \langle{T_{F1}T_{F2}T_{3_{F1}}T_{3_{F2}} \mid T_{RCN}T_{3_{RCN}}}\rangle^2
\end{equation}

To calculate the relative yield of any fragments, we multiply intensity by weight factor ($w_n$) of that particular $n$-emission channel. The weight factors are obtained by relative normalization with respect to $n$-emission channel having maximum number of counts in the corresponding partition. Therefore, the final yield of any fragment is,
\begin{equation}
I = \sum_{n} I_n \times w_n = \sum_{n} \langle{CGC}\rangle^2 \times w_n
\end{equation}
where we take summation over all the experimentally known $n$-emission channels. In the same way, we calculate the yields of all the lighter and heavier fragments in a partition. As we are not calculating the exact yields and want only relative yields, we normalize the yields of all the fragments with respect to the maximum yield. We perform this normalization separately for lighter side and heavier side of a partition. Similarly, we can have values of relative yields of fragments in all the partitions. Our calculations do not give the absolute yields at any point.

We also calculate the total mass distribution of fragments although only relative yields can be calculated. For this calculation, we use the same procedure of assigning the isospin values to fragments and then, in the same way, we calculate CGC of pair of fragments emitted in a $n$-emission channel in a partition. Since, now we are looking at complete mass distribution, we have to take into account the weightage of each partition simultaneously. For this, we normalize the weight factors ($w'_n$) of each $n$-emission channel of all the six partitions with respect to $n$-emission channel having maximum number of counts among all the partitions. Therefore, our equation for calculating the intensity of a fragment is,
\begin{equation}
I' = \sum_{n} I_n \times w'_n = \sum_{n} \langle{CGC}\rangle^2 \times w'_n
\end{equation}

\begin{figure}[h]
\centering
\includegraphics[width=9.5cm, height= 10.5cm]{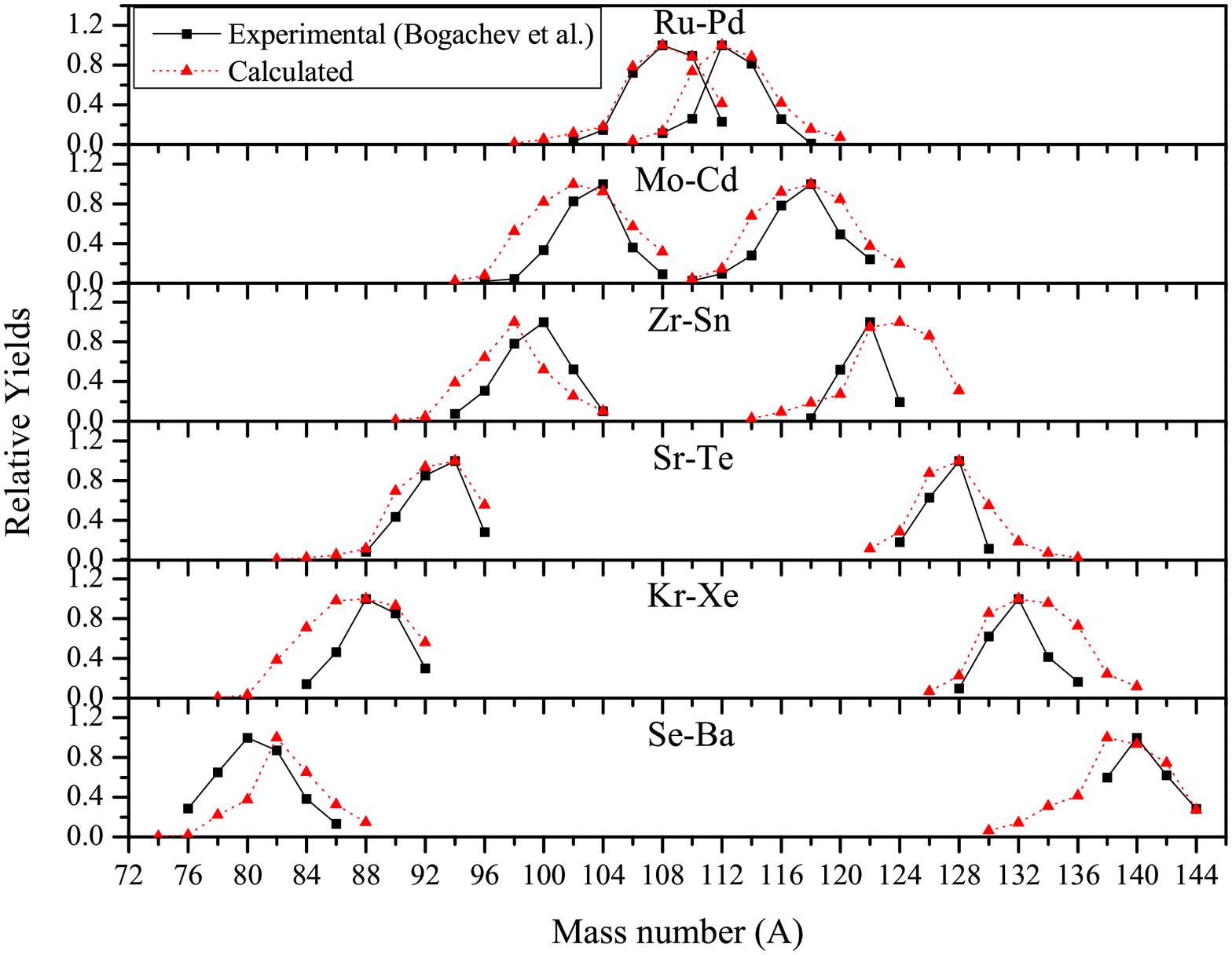}
\caption{Comparison of calculated relative yields of fragments emitted in all the six partitions with the experimental data from Bogachev et al.~\cite{bogachev}.}
\label{merge}       
\end{figure}

After calculating yields of all the fragments in individual partitions, we simply add the yields of all the three members of an isobaric multiplet to have yield corresponding to that particular mass number. Then we again normalize all the yields of mass numbers with respect to mass number with maximum yields. This gives us the relative yields of all the fragments present in the total mass distribution.
 
\section{Results and Discussion}
\label{sec-2}
We have plotted in Fig.~\ref{merge}, the calculated relative yields of fragments for each partition separately and compare them with the available experimental data of Bogachev $\it{et}$ $\it{al.}$~\cite{bogachev}. There is good agreement between the calculated and experimental data in most of the partitions. This agreement is not so good for Zr-Sn and Kr-Xe partitions. One reason for this deviation could be the presence of closed shell configuration at $A=124$ ($Z=50$) and $A=136$ ($N=82$) and their complementary fragments at $A=84$ and $A=92$~\cite{danu}. Another possibility is the presence of isomers at some of the points. Also, Bogachev $\it{et}$ $\it{al.}$~\cite{bogachev} estimated an error of 10-30\% in the experimental data. Considering these factors, we feel that our calculations reproduce the experimental trends and data reasonably well. 

\begin{figure}[h]
\centering
\includegraphics[width=9.5cm, height=8cm]{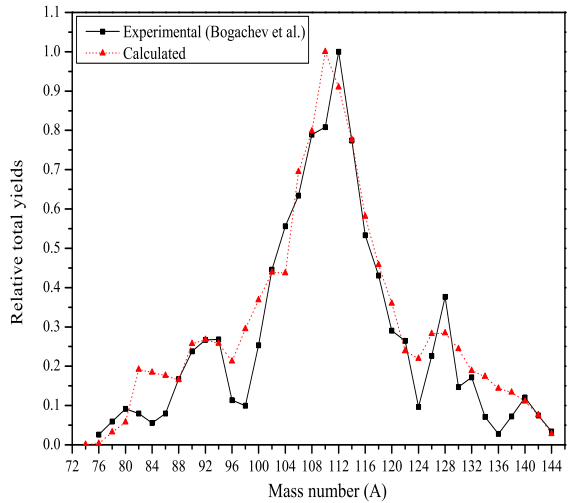}
\caption{Comparison of calculated relative total yields of fragments with the experimental data from Bogachev et al.~\cite{bogachev}.}
\label{total}       
\end{figure}

Next, we have plotted in Fig.~\ref{total} the relative total yield of fission fragments and compare it with the fission fragment mass distribution given in Bogachev $\it{et}$ $\it{al.}$~\cite{bogachev}. The two curves look very similar to each other with the exceptions at the shell closures as discussed above. These results confirm that isospin and its conservation seems to be valid in heavy neutron-rich nuclei.

\section{Conclusion}
We have calculated the relative yields of fission fragments partition-wise and also total mass distribution of fragments. In doing so we have assumed that the basic concept of isospin and its conservation remains valid in heavy nuclei. We assign the isospin values by using Kelson’s conjectures which are based on sound physics arguments. We then follow the isospin algebra and find that the fission fragments are preferably formed in isobaric analogue states forming isospin multiplets. A given isospin multiplet is assigned a $T$ value corresponding to the maximum isospin projection in the isobaric multiplet. A reasonable agreement of the calculated results with the observed fission fragment distribution provides a direct experimental evidence of the validity of isospin in heavy nuclei. Isospin, therefore, seems to emerge as a powerful tool in neutron-rich heavy nuclei and may play an important role in many phenomena and applications. 

\section*{Acknowledgment}
Support from Ministry of Human Resource Development (Government of India) to SG in the form of a fellowship is gratefully acknowledged. The authors also acknowledge the financial support in the form of travel grant from IIT Roorkee Alumni funds and IIT Roorkee.
 
%

\begin{thebibliography}{50}
%
%
\bibitem{robson}D. Robson, Science \textbf{179}, 133 (1973).
\bibitem{lane}A. M. Lane and J. M. Soper, Nucl. Phys. \textbf{37}, 663 (1962).
\bibitem{jain}A. K. Jain, D. Choudhary, and B. Maheshwari, Nuclear Data Sheets \textbf{120}, 123 (2014).
\bibitem{swati} S. Garg and A. K. Jain, Phys. Scr. \textbf{92}, 094001 (2017).
\bibitem{sliv} L. A. Sliv and Yu. I. Kharitononv, Phys. Lett. \textbf{16}, 176 (1965).
\bibitem{auerbach} N. Auerbach, Phys. Rep. \textbf{98}, 273 (1983).
\bibitem{kelson}I. Kelson, Proc. of the Conference on “Nuclear Isospin”, edited by D. Anderson, S.D. Bloom, J. Cerny and W.W. True, Academic Press (1969) Pg. 781.
\bibitem{bogachev}A. Bogachev $\it{et}$ $\it{al.}$, Eur. Phys. J. A \textbf{34}, 23 (2007).
\bibitem{danu}L. S. Danu $\it{et}$ $\it{al.}$, Phys. Rev. C \textbf{81}, 014311 (2010).


\end{thebibliography}
%
%

\end{document}